\documentclass[universe,article,accept,pdftex,moreauthors]{Definitions/mdpi}

\firstpage{1} 
\pubvolume{1}
\issuenum{1}
\articlenumber{0}
\pubyear{2024}
\copyrightyear{2024}
\externaleditor{{Academic Editor: } 
}
\datereceived{} 
\daterevised{} 
\dateaccepted{} 
\datepublished{ } 
\hreflink{https://doi.org/} 

\usepackage{graphicx,epsfig,rotate}
\usepackage[labelformat=simple]{subcaption}

\DeclareCaptionLabelFormat{subcaptionlabel}{\normalfont(\textbf{#2}\normalfont)}
\def\e10{\eta_{10}}

\def\Ombh2{\Omega_{\rm b} h^2}

\def\ga{\mathrel{\raise.3ex\hbox{$>$\kern-.75em\lower1ex\hbox{$\sim$}}}}
\def\la{\mathrel{\raise.3ex\hbox{$<$\kern-.75em\lower1ex\hbox{$\sim$}}}}

\def\beq{\begin{equation}}
\def\eeq{\end{equation}}
\def\beqar{\begin{eqnarray}}
\def\eeqar{\end{eqnarray}}

\newcommand{\trh}{T_\text{RH}}

\Title{Radiative Corrections and Reheating in Supergravity Models of Inflation \small{\rightline{UMN--TH--4537/26, FTPI--MINN--26/17, KCL–PH–TH/2026-22, CERN-TH-2026-199}}}

\TitleCitation{Radiative Corrections and Reheating in Supergravity Models of Inflation}

\Author{John Ellis$^{1,2}$, Tony Gherghetta$^3$, Kunio Kaneta$^4$, Wenqi Ke$^5$, and Keith A. Olive$^5$}

\AuthorCitation{Ellis, J.; Gherghetta, T.; Kaneta, K.; Ke, W.; Olive, K.A.}

\address{%
$^{1}$ \quad Theoretical Particle Physics and Cosmology Group, Department of
  Physics, \\   \; \ \quad King's~College~London, London WC2R 2LS, United Kingdom\\
 $^2$ \quad Theoretical Physics Department, CERN, CH-1211 Geneva 23,
  Switzerland \\
$^3$ \quad School of Physics and Astronomy, University of Minnesota, Minneapolis, MN 55455, USA\\  
$^4$ \quad Faculty of Education, Niigata University, Niigata 950-2181, Japan\\
$^{5}$ \quad William I. Fine Theoretical Physics Institute,
School of Physics and Astronomy,
University of Minnesota, Minneapolis, MN 55455, USA}

\corres{Correspondence: olive@umn.edu}

\abstract{
We consider the effects of radiative corrections in Starobinsky-like models of inflation, concentrating on models of inflation formulated in $N=1$ no-scale supergravity. 
Inflaton couplings to matter fields are necessary for reheating and these have an impact on loop corrections to the inflaton potential. Whilst corrections due to the supergravity couplings of the inflaton to Standard Model (MSSM) fields are negligible, we use {\it Planck} data to obtain interesting constraints on GUT bilinear couplings, vevs, and gauge boson masses that could be sharpened by future CMB measurements. 
}

\keyword{inflation; reheating; supergravity }

\begin{document}
\section{Introduction}

Cosmology has become a precision science, described by a Standard Model, namely $\Lambda$CDM. The driver for this theory is commonly thought to be an early epoch of near-exponential expansion, called inflation \cite{reviews}. Originally proposed as a solution to several issues in Robertson-Walker-Friedmann cosmology such as its (near-)flatness, (near-)isotropy and the absence of magnetic monopoles \cite{Guth:1980zm}, inflation has passed many observational tests. These include the generation of density perturbations by quantum effects that are approximately Gaussian and almost scale-invariant \cite{MC,pert}, as measured in observations of the cosmological microwave background (CMB) radiation \cite{Planck}. Consistency tests of inflationary models, such as higher-precision measurements of scalar perturbations, the detection of tensor (metric) perturbations in the CMB and the measurement of their spectrum, await future observations. Moreover, it is an open question how the inflaton couples to Standard Model particles and reheats the Universe following inflation.

In the mean time, every effort to improve the theoretical accuracy and precision of CMB predictions in proposed models of inflation is most welcome. As in the case of the Standard Model of particle physics, these efforts should include calculations of quantum radiative corrections to CMB observables  \cite{Gialamas:2025kef,Wolf:2025ecy,Ahmed:2025rrg,Han:2025cwk,Ellis:2025bzi,Alexandre:2025ixz,Ellis:2026ceb,Inada:2026feq,Ben-Dayan:2026bif}, including self-interactions of the inflaton and its couplings to other particles. Some may argue that inflationary models should be regarded as effective field theories that should be understood as already including radiative corrections. However, this is unsatisfactory, not least because the radiative corrections may substantially modify the form of the effective potential in a way that is not captured by simple model parametrizations. This possibility is analogous to the Standard Model Effective Field Theory (SMEFT)  approach to particle physics \cite{SMEFT1}, where radiative corrections have been shown to introduce important model sensitivities that were not apparent at the tree level \cite{SMEFT2}. Others may argue that radiative corrections are generally so small as to be negligible. This is indeed the expectation in many models of inflation in the context of supersymmetry \cite{ENOT,Nakayama:2011ri}. However, there are interesting and important exceptions, as we discuss below. In this review, dedicated to the accomplishments of Professor Ignatios Antoniadis, we revisit 
the question of radiative corrections in no-scale supergravity models of inflation. 

We first review the results from several recent papers in which different aspects of radiative corrections to inflationary models were calculated \cite{Ellis:2025bzi,Ellis:2026ceb}, with particular emphasis on Starobinsky-like models \cite{eno6} and their avatars \cite{eno7} in no-scale supergravity, and on their sensitivities to inflaton decay modes. In Section~\ref{radstar} we review the general effects of radiative corrections in Starobinsky-like models due to particles coupling to the inflaton, which are needed for reheating, deriving constraints on particle masses and the reheating temperature \cite{Ellis:2025bzi}. Then, in Section~\ref{rcns} we discuss radiative corrections arising from self-couplings of the inflaton in no-scale supergravity models \cite{Ellis:2026ceb}. Section~\ref{reheat} contains a discussion of inflaton decays in no-scale supergravity models. It is well-known that these are suppressed in wide classes of such models \cite{Endo:2006xg,egno4,Ema:2024sit,Antoniadis:2026bys}, but there are also examples where inflaton decays are not suppressed. Section~\ref{rcreh} combines results from the three previous works, considering specifically the decay channels allowed in no-scale Starobinsky-like models and computing the corrections to the inflaton potential that they induce via quantum effects. This allows us to set constraints on grand-unified-theory (GUT)-scale parameters such as the $\mu$-terms associated with the Higgs adjoint and 5-plets of SU(5). Finally, we conclude in Section~\ref{summ} with a brief summary and suggest possible directions for future studies.

\section{Radiative corrections in Starobinsky-like Models}
\label{radstar}

The Starobinsky model of inflation \cite{Staro}, originally proposed to resolve the problem of the initial big bang singularity, is formulated as an extension of Einstein gravity. Including an $R^2$ correction to the gravitational Lagrangian, a conformal transformation to the Einstein frame \cite{WhittStelle,Kalara:1990ar} results in a theory with an additional scalar field that may serve as the inflaton, $\phi$, with a potential given by
\begin{equation}
\label{treestaro}
V \; = \; \frac34 M^2 M_P^2\left[ 1 - e^{-\sqrt{\frac23} \frac{\phi}{M_P} }\right]^2 \, ,
\end{equation}
where $M$ is a mass scale and the reduced Planck mass $M_P = 1/\sqrt{8\pi G_N} = 2.435 \times 10^{18}$~GeV.
Once an inflaton potential is specified, one can compute the slow-roll parameters
\begin{equation}
\epsilon \; = \; \frac{M_P^2}{2} \left( \frac{V^\prime}{V} \right)^2 \; , \; \eta = M_P^2 \frac{V^{\prime \prime}}{V} \, ,
\end{equation}
where prime denotes a derivative with respect to $\phi$. CMB observables can be described in terms of these slow-roll parameters, as quantum perturbations. For example, the tilt of the scalar anisotropy spectrum is given by,
\begin{equation}
 n_s \; = \; 1 - 6 \epsilon_* +2 \eta_*  = \; 0.9649 \pm 0.0042 \; (68\%~{\rm C.L.})
 \, ,\label{nseq}
\end{equation}
where the numerical value is taken from the  {\it Planck} 2018 determination of $n_s$ \cite{Planck}. We note that more recent results from the ACT collaboration~\cite{AtacamaCosmologyTelescope:2025blo} may indicate a higher value of $n_s$, particularly when combined with {\it Planck}, lensing, and DESI DR2 BAO results \cite{DESI:2025zpo}:
\beq
n_s = 0.9752 \pm 0.0030 \; (68\%~{\rm C.L.}) \, .
\label{ACTns}
\eeq
The larger value of $n_s$ may call for a deformation of the Starobinsky potential \cite{Antoniadis:2025pfa,Ellis:2025ieh,EGOV} and may also arise from radiative corrections to the Starobinsky potential.~\footnote{However, recent results from the SPT collaboration~\cite{SPT-3G:2025bzu} are more consistent with those of {\it Planck}.}
The $_*$ subscript denotes field values chosen to correspond to a pivot scale, often taken to be  $k_* = 0.05$~Mpc$^{-1}$.
This is fixed by the number of $e$-folds between
the horizon exit at the pivot scale and the end of inflation: 
\begin{equation}
N_* \;\equiv\; \ln\left(\frac{a_{\rm{end}}}{a_*}\right) \; = \; \int_{t_*}^{t_{\rm{end}}} H dt \; \simeq \;  - \int^{\phi_{\rm{end}}}_{\phi_*} \frac{1}{\sqrt{2 \epsilon}} \frac{d \phi}{M_P} \, .
\label{e-folds}
\end{equation}
The end of inflation is defined when $\ddot{a}=0$, where $a$ is the cosmological scale factor. This occurs at $t_{\rm end}$, with $\phi = \phi_{\rm end}$.
Similarly $a_*$ is the value of the cosmological scale factor at the pivot scale, at $t_*$ when $\phi = \phi_*$. 
Another CMB observable is the ratio, $r$, of tensor to scalar perturbations which is given by
\beq
r \; = \; 16 \epsilon_* < 0.036 \; ,
\eeq
where the upper limit is obtained from combining \textit{Planck} data with observations by BICEP/Keck~\cite{BICEP2021,Tristram:2021tvh}.
The measured amplitude of scalar anisotropies at the pivot scale, $A^*_s = 2.1 \times 10^{-9}$ \cite{Planck} can be used to fix the inflationary mass scale, $M$, using 
\beq
A^*_s   =   \frac{V(\phi_*)}{24\pi^2 \epsilon_* M_P^4} =  \frac{3 M^2}{8\pi^2 M_P^2} \sinh^4 \left(\frac{\phi_*}{\sqrt{6}M_P} \right)\, , \label{As2}
\eeq
where the second equality is specific to the potential (\ref{treestaro}). 

All models of inflation require a mechanism to reheat the Universe and create a radiation-dominated epoch when the period of exponential expansion ends. This is most easily accomplished by inflaton decays via a coupling of the inflaton to matter \cite{dg,nos}. 
The existence of such a coupling necessarily introduces corrections to inflaton self-interactions via matter loops
that could in principle affect the inflaton potential and hence the predictions for CMB observables. 

For example, consider a possible coupling of the inflaton to a Dirac (Majorana) fermion, $y\bar{f}f\phi $ ($y\bar{f^c}f\phi/2 $). Such a coupling will result in inflaton decay, so long as $m_\phi = M > 2 m_f$. If the reheating temperature, $\trh$ is defined to be the temperature of the radiation when its energy density is equal to that in the inflaton background, we have \cite{GKMO1,Ellis:2021kad}
\beq
g_{\rm RH} \frac{\pi^2}{30} \trh^4 = \frac{12}{25} (\Gamma_\phi M_P)^2 = \frac{3y^4}{1600\pi^2}  M^2 M_P^2 \,,
\label{eq:TRH}
\eeq
where $g_{\rm RH}$ is the number of relativistic degrees of freedom at $\trh$, $\Gamma_\phi$ is the inflaton decay rate and we have assumed a decay to a Majorana fermion in the second equality.

The one-loop correction to the effective potential due to the fermion coupling is
\begin{equation}
    \Delta V _f =-\frac{n_i}{64\pi^2}\left[\mathcal{M}_f^4\left(\log \frac{|\mathcal{M}_f^2|}{\mu^2}-\frac{3}{2}\right)\right]\, ,
    \label{yukawaloop}
\end{equation}
where $n_i=2 (4)$ for a Majorana (Dirac) fermion. Note that we have not included here curvature corrections to the potential \cite{Markkanen:2018bfx}. These are typically negligible in the present context and if they dominate over the contribution in (\ref{yukawaloop}), the one-loop corrections are too small to affect the CMB observables~\cite{Ellis:2025bzi}. The inflaton field dependence in (\ref{yukawaloop}) enters through the effective mass 
\begin{equation}
    \mathcal{M}^2_f=(m_f+y\phi)^2-\frac{1}{12}R
\simeq (m_f+y\phi)^2+H^2\,.
\end{equation} 
Thus we see that the curvature corrections are sub-dominant when $y^2 \phi^2 \gg H^2$. 
For the Starobinsky potential this corresponds to $y \gg 10^{-6}$ or reheating temperatures $\trh \gg 10^8$~GeV. 

\begin{figure}[h]
    \centering
    \includegraphics[width=.48\textwidth]{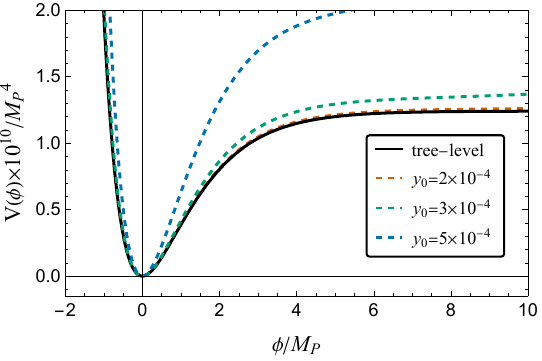}
\includegraphics[width=.48\textwidth]{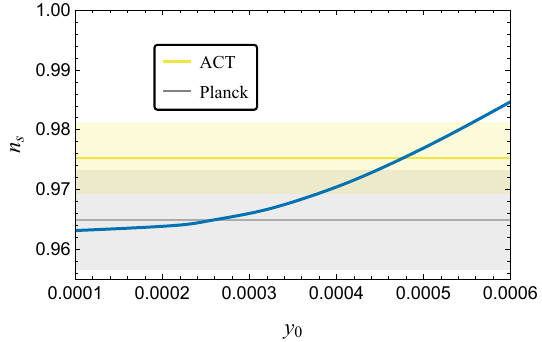}
    \caption{Left:The RG-improved effective potential as a function of $\phi$ for various values of the inflaton decay coupling $y_0$, compared to the tree-level potential. Right: The spectral index $n_s$ from the one-loop-corrected potential as a function of the decay coupling $y_0$. The gray shaded region is the {\it Planck} constraint on $n_s$ with its $\pm 2\sigma$ uncertainty, whilst the yellow band is the corresponding ACT constraint.}
    \label{efffermion}
\end{figure}

Since the couplings are renormalization-scale dependent, it is important to RG-improve the effective potential using the Callan-Symanzik equation.  Details of this calculation can be found in \cite{Ellis:2025bzi} and here we just quote the results. The one-loop RG-improved effective potential is shown in the left panel of Fig.~\ref{efffermion} for three choices of the Yukawa-like coupling, $y_0 \equiv y(\mu_0)$, where the renormalization scale $\mu_0$ has been chosen to be $M_P$ and we have assumed $m_f = 0$.  
The inflationary scale $M$ determined by Eq.~(\ref{As2}) explicitly depends on $\phi_*$ which has a weak (logarithmic) dependence on the reheating temperature and hence on $y_0$. For values of $y_0 \gtrsim 10^{-4}$, the tree-level value of $M$ is $\simeq 3.1 \times 10^{13}$~GeV, while the one-loop corrected determination of $M$
increases to roughly $4.4 \times 10^{13}$~GeV for values of $y_0\sim 0.0006$.  As one can see, for $y_0 < 2 \times 10^{-4}$ (corresponding to $\trh < 8 \times 10^{10}$~GeV, assuming $g_{\rm RH} = 427/4$) the one-loop-corrected potential is virtually indistinguishable from the tree-level Starobinsky potential. 
However, for larger values of $y_0$, the shape of the potential is affected  dramatically.

This change in the shape of the potential induces a change in the calculated value of $\phi_*$ and $M$,  and ultimately $n_s$. The change in $n_s$ as a function of $y_0$ is shown in the right panel of Fig.~\ref{efffermion}. As $y_0$ is increased above $2\times 10^{-4}$, $n_s$ increases and exceeds the 95\% upper bound from {\it Planck} data~\cite{Planck}, thereby placing an upper limit $y_0 < 4.5 \times 10^{-4}$, corresponding to $\trh < 2 \times 10^{11}$~GeV. For slightly larger $y_0$, $n_s$ is consistent with the ACT result in Eq.~(\ref{ACTns}), but the ACT results nevertheless place an upper bound $y_0 < 5.6 \times 10^{-4}$, corresponding to $\trh < 2.8 \times 10^{11}$~GeV.

\begin{figure}[h]
    \centering
\includegraphics[width=.48\textwidth]{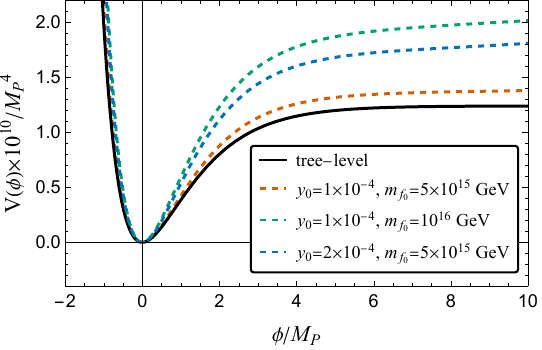}
\includegraphics[width=.48\textwidth]{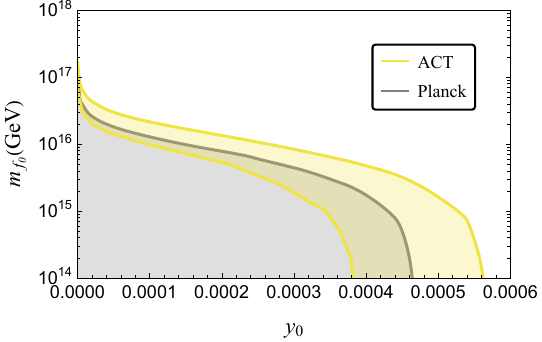}
    \caption{Left: The RG-improved effective potential as a function of $\phi$ for various values of $y_0$ and $m_{f_0}$, compared to the tree-level potential. Right: The $(y_0, m_{f_0})$ contours  satisfying the $n_s$ constraints from   \textit{Planck} (gray shaded region), and {ACT} (yellow shaded region) within $2\sigma$.
    }
    \label{efffermionmf}
\end{figure}

Similar constraints can be derived for $m_f \ne 0$ \cite{Ellis:2025bzi}, which are summarized in Fig.~\ref{efffermionmf}. The left panel again shows the effect of the radiative corrections on the potential for two choices of ${m_f}_0 = m_f(\mu_0)$ and two choices of $y_0$. The right panel shows the allowed values of ${m_f}_0$ and $y_0$ that are consistent with either the {\it Planck} or ACT results.

The above analysis assumed that the coupling of the inflaton to matter is introduced in the Einstein frame.  If a non-interacting fermion (with a kinetic term and mass term) is introduced in the $R + R^2$ (Jordan) frame, a coupling $y(\phi) = - \frac{m_f}{\sqrt{6}M_P} e^{-\frac{\phi}{\sqrt{6}M_P}}$ is generated in the Einstein frame (neglecting any coupling of the fermion to curvature). The resulting constraints on the mass $m_f$ are similar and of order $6 \times 10^{16}$ GeV. 

Limits can also be derived from coupling the inflaton to scalars. Indeed, if a scalar such as the Standard Model Higgs is introduced in the Jordan frame, a coupling of the inflaton to the Higgs kinetic term is automatically present in the Einstein frame and results in a reheating temperature of $2.9 \times 10^9$~GeV \cite{Ema:2024sit}. When a coupling such as $\frac12 \kappa \phi s^2$ between the inflaton and a scalar $s$ is introduced in the Einstein frame, we can again derive limits on the dimensionful coupling $\kappa$. In this case, for values of $\kappa \gtrsim 10^{12}$~GeV, the value of $n_s$ is decreased and {\it Planck} data set an upper bound $\kappa < 4 \times 10^{12}$~GeV, corresponding to $\trh < 4.2 \times 10^{13}$~GeV, from the {\em lower} bound on $n_s$. The ACT range for $n_s$ is never attained.

\section{Radiative corrections in no-scale supergravity models of inflation} \label{rcns}

The above constraints were derived on couplings of the inflaton to matter in the Starobinsky model in the absence of supersymmetry. However, we recall that the Starobinsky potential (\ref{treestaro}) can  be derived in the context of no-scale supergravity~\cite{eno6}. No-scale supergravity~\cite{no-scale1,EKN,ELNT,no-scale2} refers to a class of supergravity models with a maximally symmetric field-space manifold and can be characterized by the K\"ahler potential
\beq
K \; = \; -3 \, n \, \ln\left(1 - \sum _{i=1}^N\frac{|y_i|^2}{3}\right) \, ,
\label{n-sKy}
\eeq
in Planck units.
The field-space manifold has constant Ricci curvature, $R = N(N+1)/3n$ for a theory with $N$ superfields, $y_i$ and we set $n=1$ for the discussion below.

In the absence of a superpotential, the scalar potential derived from (\ref{n-sKy}) vanishes and $V = 0$.
We consider a model with two scalar fields, $y_1$ and $y_2$ where one of the fields ($y_2$) is assumed to be stabilized so that $\langle y_2 \rangle = 0$. In this case, for a non-trivial superpotential, we obtain
\begin{equation}
    V = \frac{1}{(1 - |y_1|^2/3)^2} \left( (1 - |y_1|^2/3) |W_1|^2 + |W_2|^2 - 3 |W|^2 +(y_1 W_1 W^* + {\rm h.c.}) \right)\, ,
\label{useful}
\end{equation}
where $W_{1,2} = \partial W/\partial y_{1,2}$.
We consider next a superpotential of the form \cite{Ellis:2018zya}
\beq
W(y_1, y_2)= M\left(a y_1+b y_1^2+c y_1^3+d y_2+e y_2 y_1+f y_2 y_1^2 + g(y_1, y_2)\right) \, , 
\label{genW}
\eeq
with $a,b,c,d,e,f$ arbitrary coefficients, $g(y_1, 0)
= 0$, $\partial g/ \partial y_1 (y_1 , 0)$  $= 0$ and $\partial g/ \partial y_2 (y_1, 0) = 0$.
The function $g(y_1,y_2)$ may also include terms containing factors $y_2^m$, as these would not contribute to $V$, since $\langle y_2 \rangle = 0$. 
The Starobinsky potential in Eq.~(\ref{treestaro}) is written in terms of a canonical scalar field $\phi$. This can be related to $y_1$ via
\beq
{\rm Re}~y_1 = \pm \sqrt{3} \tanh\left(\frac{\phi}{\sqrt{6}} \right)\,.
\label{branches}
\eeq
The Starobinsky potential is then given in terms of $y_1$ by 
\begin{equation}
V \; = \; \frac{M^2 |y_1|^2 ~ |1 - y_1/\sqrt{3}|^2}{(1 - |y_1|^2/3)^2} \, .
\label{V1}
\end{equation}
This potential can be obtained from (\ref{genW}) for 
\beq
a = d = 0, \qquad c = -\frac{b \left(\sqrt{1-4 b^2}+2\right)}{3 \sqrt{3}}, \qquad e = \pm \sqrt{1-4 b^2},\qquad  f = \mp \frac{\sqrt{1-4 b^2} - 2 b^2}{\sqrt{3}} \, ,
\label{solcoef}
\eeq
parameterized by $b$, with other solutions given in \cite{Ellis:2018zya}.

One specific solution is given by
$b=1/2$, $c=-1/3\sqrt{3}$, $e=0$, and $f = 1/2\sqrt{3}$, so that 
\begin{equation}
    W=M \left(\frac{ y_1^2}{2} -   \frac{y_1^3}{3 
     \sqrt{3}}  + \frac{y_1 ^2 y_2}{2 \sqrt{3}}   \right)\,.
     \label{wzybasis}
\end{equation}
By a simple field redefinition, \begin{equation}
y_1 = \frac{2 \chi}{1 + 2 T}\,, \qquad y_2 = \sqrt{3} \left( \frac{1 - 2 T}{1 + 2 T} \right) \, .
\label{Tphiwrite}
\end{equation}
the K\"ahler potential and superpotential become \cite{eno6}
\beq
K \; = \; -3  \, \ln\left(T + \bar{T} - \frac{|\chi_i|^2}{3}\right) \, ,
\label{n-sK}
\eeq
and
 \begin{equation}
W = M \left(\frac{1}{2}\chi^2 -\frac{1}{3\sqrt{3}}\chi^3\right) \,, 
\label{modelWZ}
\end{equation}
where $T$ can be associated with a modulus field and $\chi$ with a matter field.

One-loop corrections to the K\"ahler potential were derived in \cite{Gaillard:1993es,Gaillard:1996hs}. These were applied to no-scale supergravity models of inflation in \cite{Ellis:2026ceb}.
While the loop corrections to the scalar potential in supergravity involve complicated expressions, we can estimate their importance in a given model by
considering the gravitino mass and/or singularities in the K\"ahler potential. The gravitino mass is 
\beq
m_{3/2}^2 =  e^K |W|^2 \, ,
\eeq
whereas the scalar potential is
\begin{equation}
    V=F^2-3m_{3/2}^2 \quad \text{with}\quad  F=\sqrt{e^KD_iWD_{\bar{j}}\bar{W}K^{i\bar{j}}}\, ,
\end{equation}
where the $F$-term indicates the degree of supersymmetry breaking and $D_i W = W_i + K_i W$.

In the Starobinsky model, the tree-level scalar potential is flat at large field values. Thus, the supersymmetry breaking scale must increase if the gravitino mass increases at large field values.
Using Eq.~(\ref{n-sKy}) with $\langle y_2 \rangle = 0$ and Eq.~(\ref{branches}), it is easy to see that
\beq
m_{3/2} = \cosh^3(\frac{\phi}{\sqrt{6}}) |W| \, .
\eeq
To avoid an exponential increase in $m_{3/2}$, $|W|$ must vanish at large field values and we can write $W = y_2 u(y_1,y_2)$,
where $u$ is a sum of polynomials. 
But this requires the coefficients $a = b = c = 0$ in Eq.~(\ref{genW}), which is not the case for the superpotential in Eq.(\ref{wzybasis}). Thus we can expect significant radiative corrections in this case. Instead, the condition $m_{3/2} = 0$ implies that the superpotential can be written as
\begin{equation}
    W=M\left(\frac{1}{\sqrt{3}} y_1^2y_2-y_1y_2+a^\prime y_2^3-b ^\prime y_1 y_2^2+c ^\prime y_2^2\right) \, ,\label{Wm320}
\end{equation}
when restricting the couplings to at most cubic order. 

\begin{figure}[ht]
\centering\includegraphics[width=.435\textwidth]{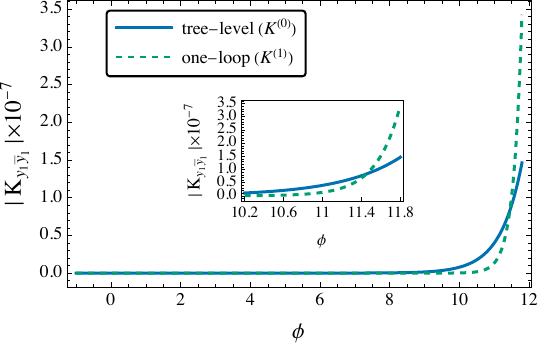} \hskip .08in
\includegraphics[width=.48\textwidth]{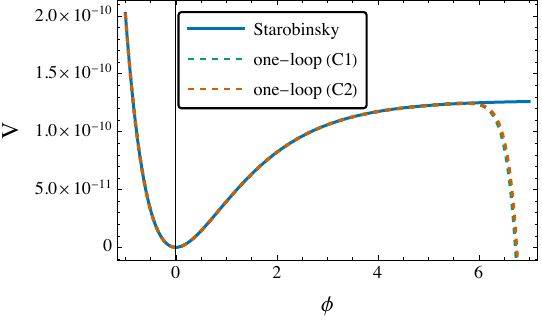}
    \caption{Left: The one-loop diagonal $y_1 {\bar y}_1$ K\"ahler metric component in the no-scale model (\ref{wzybasis}) as a function of the canonical scalar field $\phi$.  We fix $\mu=M=1.3\times 10^{-5}$ and $\lambda_i(\mu_0)=1$, $\mu_0 = \Lambda=0.009$, $c^\prime=0.01$ (C1). Right: The one-loop corrected potential compared with the Starobinsky potential \eqref{treestaro} as a function of the canonical scalar field $\phi$. Green dashed curve: one-loop corrected potential with the first set of conditions (C1); red dashed curve: one-loop corrected potential with the second set of conditions (C2).}
    \label{kcompWZ}
\end{figure}

Singularities in the K\"ahler potential may also signal sizeable radiative corrections. While the K\"ahler potential itself is logarithmically singular as $y_1 \to \sqrt{3}$ (or $\phi \to \infty$), the field space metric may diverge as $1/(y_1-\sqrt{3})^k$ with $k=2$ and the scalar potential may possess an even stronger divergence ($k>2$). In this case,
the one-loop corrections may dominate at large $\phi$ and deform the potential. Interestingly, the condition $m_{3/2} = 0$ with superpotential (\ref{Wm320}) ensures the cancellation of $k=4$ and $k=5$ terms in the potential. The requirement that $k=3$ or there be no singularity in the one-loop K\"ahler correction is satisfied for the choice $c' = \sqrt{3} b'$.
However, neither of these conditions is satisfied for the model described by (\ref{wzybasis}), and the potential is deformed at large $\phi$ as can be seen in Fig.~\ref{kcompWZ}, where we show the tree-level and one-loop corrected K\"ahler metric component $K_{y_1 {\bar y}_1}$ in the left panel.

To ensure the stabilization of $y_2$, we may introduce a quartic term, $|y_2|^4/\Lambda^2$, in the logarithm of (\ref{n-sKy}) \cite{Ellis:1984bs,eno7,egno4,building} where $\Lambda$ is a UV scale, as well as a quadratic term in the superpotential, $ c' y_2^2$,  which does not affect the potential along the inflationary direction ($y_2 = 0$). 
To compute the radiative corrections, we allow the three coefficients in (\ref{wzybasis}) to run by introducing couplings $\lambda_i$ with $\lambda_i (\mu_0) = 1$ for $i=1,2,3$, $\mu_0 = \Lambda = 0.009 M_P$ and $c' = 0.01$ (C1).
As one can see in Fig.~\ref{kcompWZ} the tree-level and one-loop metric components start to deviate at around $\phi \simeq 10 M_P$. 

In the right panel of Fig.~\ref{kcompWZ}, we show the Starobinsky potential compared to the one-loop corrected scalar potential for two choices of initial parameter values: C1 as defined above and C2 with  $\Lambda=0.006 M_P$,  $c^\prime (\mu_0)=0.01$, $\lambda_1(\mu_0)=1 + 10^{-5},\lambda_2(\mu_0)=1 +3\times  10^{-5},\lambda_3(\mu_0)=1$, and fixing the initial scale $\mu_0=\Lambda$. We see that the radiative corrections are indeed significant for $\phi \gtrsim 6 M_P$. At this field value,  the  gravitino mass is of order $100M$, so the supersymmetry breaking scale $\sqrt{F} \sim \sqrt{m_{3/2} M_P} > \Lambda$, indicating a breakdown of our effective treatment of stabilization \cite{Dudas:2017kfz}.

\begin{figure}[ht]
\centering\includegraphics[width=.43\textwidth]{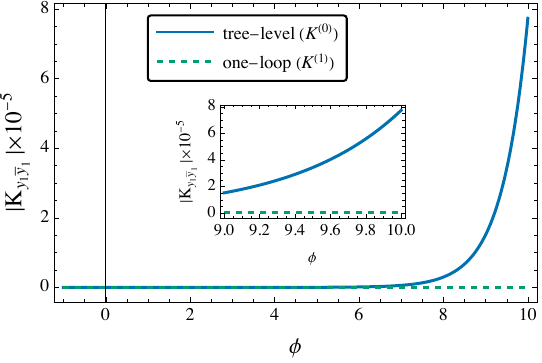} \hskip .08in \includegraphics[width=.48\textwidth]{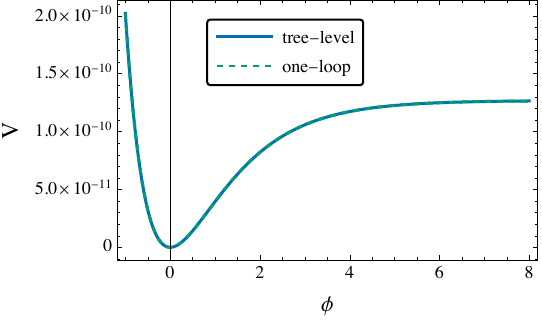}
    \caption{Left: Comparison of the tree-level and one-loop contributions to the $y_1 {\bar y}_1$ component of the K\"ahler metric in the Cecotti model~\cite{Cecotti}, as functions of the canonically-normalized inflaton field $\phi$.  Right: Comparison of the tree-level scalar potential and its one-loop corrected form in the Cecotti model~\cite{Cecotti}.}
    \label{kcompc}
\end{figure}

In contrast, another way to obtain the Starobinsky potential from Eq.~(\ref{branches}) is by taking 
$b=0$, which implies $c=0$, $e=-1$ and $f=1/\sqrt{3}$, giving the superpotential
\beq
W=M\left(-1+\frac{1}{\sqrt{3}} y_1\right)y_1 y_2 \, .
\label{Cy1y2}
\eeq
Note that this is precisely of the form needed to insure that the gravitino mass vanishes at large field values. That is, it is given by Eq.~(\ref{Wm320}) with $a' = b' = c' = 0$ and is guaranteed not to exhibit dangerous divergences. After a transformation $y_1 \to -y_2$ and $y_2 \to - y_1$ and transformation to the $T, \chi$ basis, one recovers the familiar form of the Cecotti model \cite{Cecotti}
\beq
W=\sqrt{3} M \chi \left(T - \frac12\right)
\, .
\label{WTchiC}
\eeq
The canonical inflaton is related to $T$ through
\beq
T = \frac12 e^{\sqrt{\frac23} \phi} \, .
\label{canT}
\eeq
We also define the canonically normalized fluctuation in T by
\begin{equation}
T+\bar T
\;=\;
\exp\!\left[
\sqrt{\frac{2}{3}}\,\delta T
\right]\,,
\qquad
\langle T+\bar T\rangle=1\,.
\label{candeltaT}
\end{equation}
To compute the loop correction, we  again introduce running couplings $\lambda_1(\mu)$ and $\lambda_2(\mu)$ and evaluate at the scale $\mu=M_0=1.3\times 10^{-5}$. We choose 
$\lambda_i(\mu_0)=1$, $M(\mu_0)=M_0$ with $\mu_0=\Lambda = M_P$. The scale $\Lambda$ is again a quartic correction to insure stabilization in the $y_2$ direction. 
The tree-level and one-loop contributions to the $y_1 {\bar y}_1$ component of the K\"ahler metric evaluated with $y_2 = 0$ are shown in the left panel of Fig.~\ref{kcompc}. We see that the one-loop correction in this case is always much smaller than the tree-level component, and we also see in the right panel of Fig.~\ref{kcompc} that there is very little difference between the tree-level and one-loop potential for the Cecotti model.

\section{Reheating in supergravity models of inflation}
\label{reheat}

As noted earlier, reheating is an essential aspect of any inflationary model, and may be achieved directly through inflaton decay if there are suitable couplings between the inflaton and Standard Model fields. We noted in Section~\ref{radstar} that the Starobinsky model, when formulated in the Jordan frame, naturally contains a decay channel to the Higgs boson arising from its coupling to the Higgs kinetic term, leading to a reheating temperature of $\trh = 2.9 \times 10^9$~GeV.  This result is modified if the Higgs is coupled to curvature, $\xi R |H|^2$, as the inflaton decay rate becomes proportional to $(1 - 6 \xi)^2$  \cite{Watanabe:2006ku,Ema:2024sit}. Thus, this rate vanishes for $\xi = 1/6$. 
However, the Starobinsky inflaton may still have an appreciable decay rate, even if $\xi = 1/6$, through its coupling to the trace anomaly \cite{Gorbunov:2012ns} that generates decays to gauge bosons \cite{Kamada:2019pmx}, leading to a reheating temperature of order $10^8$~GeV.

That the Starobinsky model can be derived from no-scale supergravity (as described in the previous Section) or from an $R+R^2$ theory of gravity indicates a deeper correspondence between the theories, as has been explored in \cite{eno9,DLT, building,Antoniadis:2026uzn,Antoniadis:2026bys}. For example, one can relate the conformal factor ($e^{2\Omega}$) between the Einstein and Jordan frames of the $R+R^2$ theory and the K\"ahler potential through 
\beq
\Omega = - \frac16 K \, .
\eeq
In addition, Standard Model scalar fields, such as Higgs fields appearing as untwisted fields in the K\"ahler potential (cf., the $y_i$ in Eq.~(\ref{n-sKy})) that are not related to the inflationary sector) can be matched to conformally-coupled scalar fields with $\xi = 1/6$ in the Jordan form of the $R+R^2$ theory.  Thus the coupling of the inflaton to the Higgs kinetic term is expected to vanish, leaving only a coupling to the mass term in the Higgs potential (which breaks the conformal invariance) so that the inflaton decay rate is proportional to $m_H^4$, which is very small in the Standard Model.
Indeed this is what is found when one expands the effective Lagrangian to determine the linear couplings of the inflaton, as discussed in more detail in \cite{egno4,building,Antoniadis:2026bys}.  

In the remainder of this Section, we consider only the decays of the inflaton candidate in the Cecotti model defined by the superpotential in Eq.~(\ref{WTchiC}). 
For example, the decay rate of the inflaton to two scalars ($\Phi_I, {\bar \Phi}^J$) is given by \cite{egno4,Antoniadis:2026bys}
\begin{equation}
\Gamma\bigl(\delta T \to \Phi_I\bar\Phi^J\bigr)
\;=\;
\frac{\bigl(n_I+n_L-3n\bigr)^{2}}{n}
\frac{\bigl|W^{IL}\bar W_{LJ}\bigr|^{2}}{48\pi\,M\,M_{P}^{2}}\,,
\label{eq:Gamma-2body-scalar}
\end{equation}
where $\delta T$ is the fluctuation associated with the canonically normalized inflaton given in Eq.~(\ref{candeltaT}), 
$n$, is the coefficient of the logarithm
in the K\"ahler potential as defined in Eq.~(\ref{n-sKy}), $W_I = \partial W/\partial \Phi_I$, and similarly for other derivatives of the superpotential. The parameters $n_I$ correspond to modular weights
of the scalars. For untwisted scalars (fields inside the logarithm in $K$), $n_I =1$, but we allow for the presence of twisted fields whose kinetic terms are outside the logarithm, with $K \supset |\varphi_a|^2/(T+{\bar T})^{n_a}$, possibly with $n_a \ne 1$.
The only non-vanishing contribution to the decay rate in Eq.~(\ref{eq:Gamma-2body-scalar}) in the minimal supersymmetric Standard Model (MSSM) comes from the Higgs bilinear, $W\supset \mu_H H_u H_d$, so that the rate is proportional to $\mu_H^4/M M_P^2$, where $M$ is the inflaton mass. This rate is negligible in the MSSM, but can be important in models of high-scale supersymmetry \cite{Dudas:2017kfz,Kaneta:2019yjn}. This channel may also be important if there exist massive scalars, associated with an intermediate or grand unified scale and having a bilinear mass term for scalars with masses close to the inflaton mass, $M\simeq 3 \times 10^{13}$~GeV. 
There are also 3- and 4-body channels for decays to scalars. But these either lead to low ($\sim$ MeV) reheating temperatures (in the case of 3-body decays, or vanish when $n=1$ and the modular weights $n_a =1$ (in the case of 4-body decays). For $n\ne 1$ or $n_a \ne 1$, this channel can lead to reheating temperatures of order $10^7$~GeV.  However, as these channels do not make relevant contributions to the radiatively-corrected scalar potential, we do not consider them further here. 

Similarly, the rate of inflaton decay to two fermions, (${\bar \chi}_I, \chi_J$) is given by
\begin{equation}
\Gamma\bigl(\delta T \to \bar\chi_I\chi_J\bigr)
\;=\;
\frac{(n_I+n_J-3n)^{2}}{n}\,
\frac{|W^{IJ}|^{2}\,M}{192\pi\,M_{P}^{2}}\,.
\label{eq:Gamma-2body-fermion}
\end{equation}
This rate is significantly larger than the two-body rate to scalars but still only provides reheating up to temperatures of order 10 MeV for MSSM fermions. This coupling will also provide a negligible contribution to the one-loop inflaton potential unless the fermions are associated with an intermediate scale and have a bilinear coupling, $\mu_I \gtrsim M$, a possibility that is discussed further in the next Section.
Three-body decay channels involving the Higgs and two matter fermions can lead to reheating temperatures of order $10^8$~GeV, if either $n\ne 1$ or some field has modular weight different from 1. 

We noted that in the Starobinsky model, the inflaton naturally couples to the trace of the energy momentum tensor  \cite{Gorbunov:2012ns}
\begin{align}
	\mathcal{L} \; \supset \; \Omega {T^\mu}_\mu = \frac{\phi}{\sqrt{6}\,M_P}
{T^\mu}_\mu \, ,
 \label{OT}
\end{align}
which in supergravity would  correspond to a coupling of the K\"ahler potential to the trace anomaly \cite{Endo:2007sz,anom}
\begin{align}
	\mathcal{L} = -\frac{K}{6} {T^\mu}_\mu \, .
 \label{KT}
\end{align}
This could lead to decays of the inflaton into pairs of gauge bosons. However, such a conclusion contains some subtleties which we now review \cite{Antoniadis:2026bys}.
First let us suppose that the gauge kinetic function, $f_{\alpha \beta}$, defined by
\begin{align}
{\cal L}_{G}
&\supset
-\frac{1}{4}
\bigl({\rm Re}\,f_{i\alpha\beta}\bigr)
F^{\alpha}_{i\mu\nu}F_i^{\beta\,\mu\nu}
\, ,
\label{eq:LG}
\end{align}
is non-minimal for gauge group $i$ and
depends explicitly on $T$, and hence on the inflaton, $f_{i\alpha\beta} = f_i(T) \delta_{\alpha \beta}$.  The inflaton-gauge coupling is then proportional to \cite{Endo:2006xg,egno4}
\begin{equation}
d_{i,T}
\equiv
\langle{\rm Re}\,f_i\rangle^{-1}
\left|
\left.
\frac{\partial f_i}{\partial T}
\right|_0\right|\, ,
\label{eq:dgT-def}
\end{equation} 
where $\partial f_i/\partial T$ is evaluated at the minimum of the scalar potential with $\langle T+\bar T\rangle=1$.  The partial width for inflaton decay to 
gauge bosons is then
\begin{equation}
\Gamma(\delta T\rightarrow A_iA_i)
=
\frac{d_{i,T}^{\,2}}{32\pi}
\left(\frac{N_i}{12}\right)
\frac{M^3}{M_P^2} \, ,
\label{eq:Gamma-direct}
\end{equation}
where $N_i=\dim G_i$ is the number of gauge bosons in the final-state
sector and $M$ is the inflaton mass. 
Note that the partial width for the decay to gauginos is suppressed by a factor of $(m_{3/2}/M)^2$ relative to (\ref{eq:Gamma-direct}) \cite{Kallosh:2011qk}.

In the case where the gauge kinetic function is minimal, ie., 
$f_{i\alpha\beta} = \delta_{\alpha \beta}$, there is no direct tree-level coupling of the inflaton to gauge bosons. However, via the trace anomaly, there may still be a coupling, since 
\begin{align}
	\left.{T^\mu}_\mu\right\vert_\mathrm{anom} = \sum_{i,\alpha} \frac{\beta_i}{4\alpha_i} F^{\alpha}_{i\mu\nu} F_i^{\alpha \mu\nu}\,,
 \label{anomT}
\end{align}
where $i$ runs over all the different gauge groups, $\alpha$ runs over the generators of each gauge group,
and $\alpha_i$ is the fine structure constant of gauge group $i$.
The beta function is given~by
\begin{align}
	\beta_i = \frac{b_i \alpha_i^2}{2\pi}\,,
\end{align}
where $b_i$ counts only the number of light degrees of freedom as explained above, with
$b_1 = 33/5$ for U(1), $b_2 = 1$ for SU(2), and~$b_3 = -3$ for SU(3) in the MSSM.

We define the one-loop inverse gauge coupling for the no-scale model under consideration by
\begin{equation}
{\cal F}_i
=
{\rm Re}\,f_i
+
\frac{b_i}{16\pi^2}
\log (T+\bar{T}) \, ,
\label{eq:FaT}
\end{equation}
in the $(T,\chi)$ basis,
or
\begin{equation}
{\cal F}_i
=
{\rm Re}\,f_i
+
\frac{b_i}{16\pi^2}
\log \left(1-\frac13|y_1|^2\right) \, ,
\label{eq:Fay}
\end{equation}
in the $y_i$ basis. 
The coupling of the inflaton to the gauge bosons is then proportional to 
\beq
g_i^2 \left.\left|\frac{\partial{\cal F}_i}{\partial \delta  T}\right|_0 \right| = 
g_i^2 \left.\left|\left(\frac{\partial{\cal F}_i}{\partial T}+\frac{\partial{\cal F}_i}{\partial \bar T}\right)\right|_0 \right| \, \frac{\partial T}{\partial \delta T} \, ,
\eeq
in the $(T, \chi)$ basis, with a similar expression in the $y_i$ basis. 

If $f_{i\alpha\beta}$ is minimal in the case of the $y_i$ basis with the K\"ahler potential defined by Eq.~(\ref{n-sKy}), there is still no direct coupling of the inflaton to gauge bosons, as the linear coupling of the inflaton at the minimum of the potential at $y_1 = 0$, $\partial {\cal F}_i/\partial \delta y$ vanishes. However, if $f_{\alpha\beta}$ is minimal in the $(T,\chi)$ basis with the K\"ahler potential defined by Eq.~(\ref{n-sK}), then since $\partial {\cal F}_i/\partial \delta T \ne 0$,
there is a coupling 
\begin{equation}
{\cal L}_{\delta TFF}
=
-\frac{1}{4}
\frac{d_{a,T}^{\rm anom}}{\sqrt{6}}
\frac{\delta T}{M_P}\,
F^a_{\mu\nu}F^{a\,\mu\nu}  \, ,
\label{eq:L-eff-from-F}
\end{equation}
with
\begin{equation}
d_{a,T}^{\rm anom}
= \; \frac{\alpha_i}{2\pi} |b_i| \, ,
\label{eq:d-anom-final}
\end{equation}
where we have assumed all modular weights, $n_I = 1$ and $n=1$.
This leads to a decay rate
\begin{equation}
\Gamma(\delta T\to A_aA_a)\big|_{\rm anom}
=
\frac{N_a \alpha_i^2}{1536\pi^3}
\left|b_i\right|^2
\frac{M^3}{M_P^2} \, .
\label{eq:Gamma-anom}
\end{equation}
leading to a reheating temperature of 
\begin{equation}
\trh
=
\left(
\frac{72}{5\pi^2g_{\rm RH}}
\right)^{1/4}
\left( \sum_a \frac{N_a b_a^2 \alpha_a^2}{1536 \pi^{3}}\right)^{1/2} \frac{M^{3/2}}{M_P^{1/2}} \simeq 4 \times 10^7~{\rm GeV} \, ,
\label{eq:Treh-Gamma2}
\end{equation}
where we have taken $M=3 \times 10^{13}$~GeV.

One may ask at this point how can a change in basis result in zero coupling (in the $y_i$ basis) and a decay rate given by Eq.~(\ref{eq:Gamma-anom}) in the $(T,\chi)$ basis? The answer lies in a careful determination of the K\"ahler transformation between the two bases \cite{Kaplunovsky:1994fg,Antoniadis:2026bys}. Indeed it is not possible to start with the tree-level gauge kinetic function being minimal in both bases. 
If the gauge kinetic function is minimal in the $y_i$ basis, the coupling of the inflaton to gauge bosons vanishes. 
For the transformation in Eq.~(\ref{Tphiwrite}), the K\"ahler potential transforms as 
\beq
K(y_1,y_2) \to K (T,\chi) + 3 \log \left|T+\frac12\right|^2 \, ,
\eeq
and the gauge kinetic function transforms as 
\beq
f_{\alpha\beta} \to f_{\alpha\beta} - \frac{b_i \delta_{\alpha\beta}}{16\pi^2}
\log (T+\bar{T}) \, .
\label{kftrans}
\eeq
The second term on the right hand side of Eq.~(\ref{kftrans}) exactly cancels the contribution arising from the trace anomaly and the net coupling of the inflaton to gauge bosons vanishes. If on the other hand, the tree-level gauge kinetic function is minimal in the $(T,\chi)$ basis, the coupling is given by Eq.~(\ref{eq:L-eff-from-F}). In the $y_i$ basis, although the trace anomaly contribution vanishes, 
the tree level kinetic kinetic function becomes 
\beq
f_{\alpha\beta} \to f_{\alpha\beta} - \frac{b_i}{16\pi^2}
\log \left|1+\frac{1}{\sqrt{3}}y_1\right|^2 \, ,
\eeq
and leads to the identical coupling given in Eq.~(\ref{eq:L-eff-from-F}). For further details and more general expressions for these couplings, see \cite{Antoniadis:2026bys}.

\section{Radiative correction from supergravity couplings and reheating.}
\label{rcreh}

In this section, we combine our calculations of decay couplings of the inflaton to matter fields leading with the results reviewed in Section \ref{radstar}. In particular, we will derive bounds on GUT scale parameters such as the bilinear terms associtated with the SU(5) adjoint, and 5-plets, the adjoint vev, and gauge boson masses.

We begin by considering the corrections to the scalar potential from scalar loops induced by the coupling of the inflaton to matter scalars. The decay rate given in Eq.~(\ref{eq:Gamma-2body-scalar}) is derived from the coupling \cite{egno4,Antoniadis:2026bys}
\beq
\mathcal{L}
\supset
-\frac{\delta T}{\sqrt{3n}}\,
\left(n_I+n_L-3n\right)
W^{IL}\bar W_{LJ}\Phi_I\bar\Phi^J  = \frac{\mu_H^2}{\sqrt{3} M_P} \delta T (|H_u|^2 + |H_d|^2) \, ,
\label{scalardec}
\eeq
where in the second equality we have taken $n=n_I=n_L = 1$, and $H_u,H_d$ are the MSSM Higgs doublets. 
In addition, there are loop contributions from potential 3-body decay channels, when one of the scalars is a Higgs boson with a non-zero vacuum expectation value (vev), $v$. However, we do not consider them here, as they are suppressed by a factor $v/\mu_H$. 
The Lagrangian term in (\ref{scalardec})
corresponds to a coupling 
\beq
\kappa = \frac{\mu_H^2}{\sqrt{3}M_P} \, ,
\eeq
for each of the $N_H = 8$ real scalars in the MSSM present in the loop correction.

 In the case of a weak scale value for $\mu_H$, $\kappa$ is very small and makes a negligible contribution to the inflaton potential.  
However, in any GUT,
we expect GUT scale Higgs bosons with a corresponding $\mu$ parameter. For example, in a minimal supersymmetric SU(5) model we expect the superpotential to contain the following terms:
\beq
W
\supset \mu_\Sigma {\rm Tr}\Sigma^2 + \frac{1}{6} \lambda^\prime {\rm
 Tr} \Sigma^3 + \mu_H \overline{H} H + \lambda \overline{H} \Sigma H \, ,
\eeq
where $\Sigma$ is the SU(5) adjoint and $H$ ($\overline{H}$)  are the SU(5) Higgs 5-plets containing the MSSM $H_u$ ($H_d$) Higgs doublets. The vev of $\Sigma$ is $v_\Sigma= 4 \mu_\Sigma/\lambda'$, and the masses of adjoint and SU(5) superheavy gauge bosons are given by $M_\Sigma = \frac52 \lambda' v_\Sigma= 10 \mu_\Sigma$ and $M_X = 5 g_5 v_\Sigma= 20 g_5 \mu_\Sigma/\lambda'$, where $g_5$ is the SU(5) gauge coupling. GUT matching conditions can be used to determine $g_5$
and the combination $M_X^2 M_\Sigma$. Therefore, for a given value of $\lambda'$,
$v_\Sigma$ and $\mu_\Sigma$ can be determined. 
For more details see e.g., \cite{Ellis:2016tjc}.

The limit \cite{Ellis:2025bzi} $\kappa < 4\times 10^{12}$~GeV can be translated into a limit on $\mu_\Sigma$
\beq
\mu_\Sigma \lesssim 4.1 \times 10^{15}~{\rm GeV} \frac{1}{N_H^\frac14}\,,
\label{S1}
\eeq
and correspondingly a limit on the adjoint Higgs vev
\beq
v_\Sigma \lesssim 1.6 \times 10^{16}~{\rm GeV} \frac{1}{\lambda' N_H^\frac14} \, ,
\label{S2}
\eeq
where $N_H = 48$ is the number of real scalars in the complex adjoint.  We note that in many phenomenological models \cite{Ellis:2016tjc}, proton decay limits typically require a large vev and hence a small value of $\lambda' \sim 10^{-4}$.
However this requirement is relaxed when
non-renormalizable operators play a role
allowing $\lambda' \sim 1$
\cite{Ellis:2026tgb}, in which case the limits in (\ref{S1}) and (\ref{S2}) are non-trivial. 

Similar bounds can be derived from GUT fermion loops. The decay rate in Eq.~(\ref{eq:Gamma-2body-fermion}) is derived from \cite{egno4,Antoniadis:2026bys}
\beq
{\cal L} \supset
-\,\frac{\delta T}{2\sqrt{3n}}
\bigl(n_I+n_J-3n\bigr)\,W^{IJ}\,
\bar\chi_{I,L}\chi_{J,R} = \frac{\mu_H}{2 \sqrt{3} M_P} \delta T{\widetilde H_u} {\widetilde H_d} \, ,
\label{fermdec}
\eeq
where again in the second equality we have taken $n=n_I=n_L = 1$, and ${\widetilde H_u}$ and  ${\widetilde H_d}$ are the Higgsinos associated with the MSSM Higgs doublets. There are also 3-body decay channels involving a Higgs and fermion pair but, as these are also suppressed by a factor $v/\mu_H$, we neglect them here. The coupling in (\ref{fermdec}) corresponds to a Yukawa-like coupling
\beq
y_0 = \frac{\mu_H}{2 \sqrt{3} M_P} \, ,
\eeq
for each of the $N_{\tilde H} = 2$ Higgsinos in the MSSM. 
As in the scalar case, this is a very small Yukawa coupling in the MSSM with weak scale supersymmetry breaking. However, the limit $y_0 < 4.5 \times 10^{-4}$ results in a limit on the GUT-scale $\mu$-term
\beq
\mu_\Sigma \lesssim 3.8 \times 10^{15}~{\rm GeV} \frac{1}{\sqrt{N_{\widetilde H}}} \, ,
\eeq
and provides a comparable limit on the GUT scale parameters as in the scalar case. Note that while the coupling in Eq.~(\ref{fermdec}) contributes to inflaton decay for the light Higgsinos, it is also relevant for GUT scale Higgsinos due to the loop corrections to the potential though inflaton decays to GUT scale Higgsinos are kinematically forbidden. 

Finally, we consider the effects of gauge boson loops. 
For a massive gauge boson with a mass term $m_V^2 V_\mu V^\mu$ and coupling to the inflaton through the gauge kinetic function $\mathcal{F}_i(T)$  in Eq.~\eqref{eq:FaT}, we can make a field redefinition $V_\mu^\prime\equiv \sqrt{\mathcal{F}_i(T)}V_\mu$ so that the kinetic term becomes canonical. In this case, the field-dependent mass is $m_V^\prime (T)=\frac{m_V}{\sqrt{\mathcal{F}_i(T)}}$. The one-loop Coleman-Weinberg potential is
\begin{equation}
    \Delta V=\frac{3 N_V}{64\pi^2} m_V^{\prime 4}(T)\left(\log \frac{m_V^{\prime 2}(T)}{\mu^2}-\frac{5}{6}\right)
    \label{gaugeboson}
\end{equation}
where $N_V$ is the number of massive gauge bosons. 

We note that if supersymmetry is broken at a scale much below the GUT scale, one must also take into account the contribution from gaugino loops, which enter into $\Delta V$ with the opposite sign \cite{ENOT,Nakayama:2011ri}. Instead of $\Delta V \propto m_V^{\prime 4}$, we would expect $\Delta V \propto m_{3/2}^2 m_V^{\prime 2}$. 

We argued in Section \ref{rcns} that supersymmetry breaking is small in the Cecotti model (\ref{WTchiC}). In this case, we would expect the corrections from Eq.~(\ref{gaugeboson}) to be negligible.  However, a small gravitino mass is not inevitable in no-scale supergravity avatars of Starobinsky-like inflation \cite{eno7,Ellis:2018zya}, so it is of interest to determine the constraints that can be obtained from Eq.~(\ref{gaugeboson}). 
As an example, we consider minimal $f_{i\alpha\beta}$, in which case the coupling to the inflaton is generated by the trace anomaly alone. In terms of the inflaton $\phi$, the field-dependent mass is 
\begin{equation}
    m_V^{\prime 2} (\phi )=\frac{m_V^2}{1+\frac{b_i}{16\pi^2}\sqrt{\frac{2}{3}}\frac{\phi}{M_P}}\label{mvphi} \, .
\end{equation}
Therefore, the one-loop correction is larger when $b_i < 0$  for larger field values, and becomes important when $ m_V^{\prime 2} \sim M M_P$. Assuming that the denominator in \eqref{mvphi} is $\mathcal{O}(1)$, the one-loop correction is large when  $m_V\approx \mathcal{O}(10^{16})$~GeV.~\footnote{We do not consider here two-loop contributions, which could become important under some circumstances.} 

\begin{figure}[ht!]
\centering\includegraphics[width=.5\textwidth]{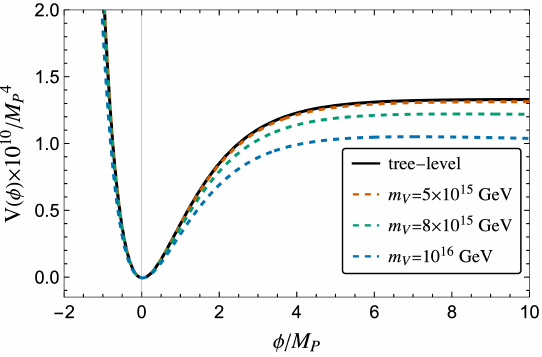} \hskip .08in
    \caption{Comparison of the tree-level and one-loop scalar potentials for different values of the bare mass $m_V$. We consider  $SU(5)$ and $\trh=10^8\text{ GeV}$.}
    \label{nsrboson}
\end{figure}

\begin{figure}[ht!]
\centering\includegraphics[width=.5\textwidth]{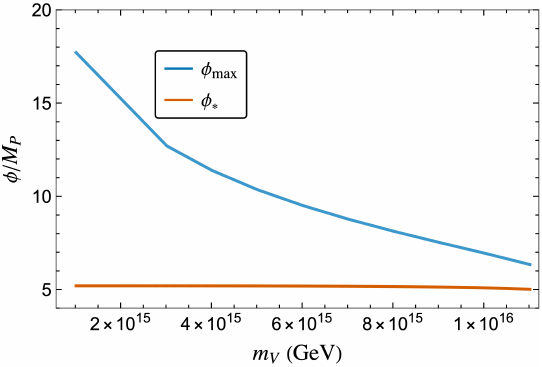} \hskip .08in
    \caption{The values of $\phi_\text{max}$, where the one-loop potential
has a local maximum, and $\phi_*$ as functions of the bare mass $m_V$. Here we fix $\trh=10^8\text{ GeV}$.}
    \label{phimax}
\end{figure}

To illustrate the constraint on the bare mass $m_V$, we fix    $\mu^2=m_V^{\prime 2}(\phi)$ and consider $SU(5)$, for which $b_5=-3$ and $N_V=12$ is the number of massive SU(5) gauge bosons after symmetry breaking.
We further assume $\trh=10^8$~GeV for this illustration. In Fig.~\ref{nsrboson},
we compare the one-loop corrected scalar potential to the tree-level potential for three values of the gauge boson mass. Though it is not seen in the figure, the radiative corrections cause the slope of the potential to turn negative at large values of $\phi/M_P$.
The maximum of the potential occurs at $\phi_{\rm max}$, which is plotted versus the gauge boson mass in Fig.~\ref{phimax}. 
As one can see, $\phi_{\rm max}$ drops from 18$M_P$ to about 6 $M_P$ as $m_V$ varies from $2 \times 10^{15}$~GeV to $1.1 \times 10^{16}$~GeV. This is similar to what was found when considering the contributions of scalar loops and large couplings $\kappa \approx \mathcal{O}(10^{12}) $~GeV \cite{Ellis:2026ceb}.

\begin{figure}[ht]
\centering\includegraphics[width=.5\textwidth]{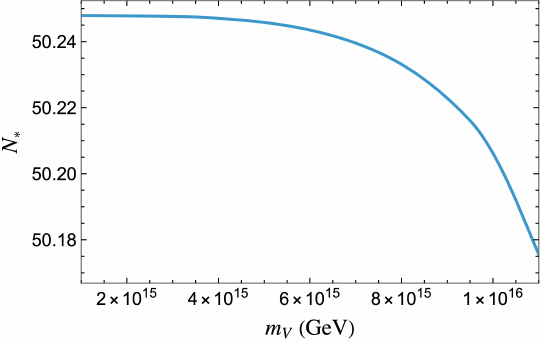} \hskip .08in
    \caption{The number of e-folds $N_*$ obtained from the one-loop
corrected potential as a function of the gauge boson mass.}
    \label{nsrbosonnstar}
\end{figure}

The number of e-folds, $N_*$ varies only slightly as a function of the gauge boson, as is shown in Fig.~\ref{nsrbosonnstar}. The corresponding value of $\phi_*$ is also nearly constant, as is shown by the red curve in Fig.~\ref{phimax}.

\begin{figure}[ht]
\centering\includegraphics[width=.45\textwidth]{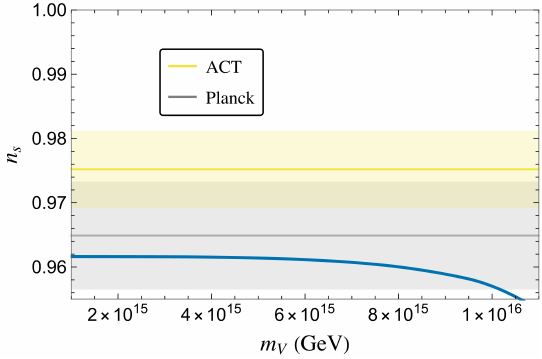} \hskip .08in
\includegraphics[width=.45\textwidth]{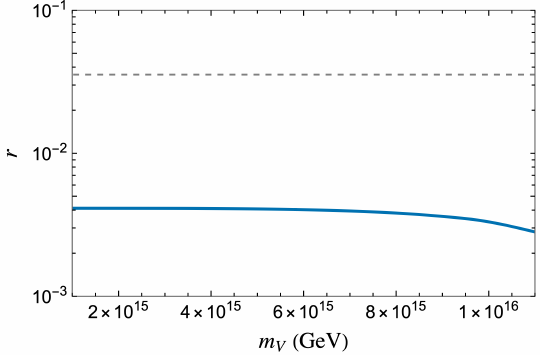}
    \caption{Left panel: Comparison of $n_s$ as a function of $m_V$ with the {\it Planck} and ACT bounds. Right panel: Comparison of $r$ as a function of $m_V$ with the {\it Planck} upper bound (gray dashed line). Here we choose $\trh=10^8\text{ GeV}$.}
    \label{nsrbosonnsr}
\end{figure}

Finally, given the value of $\phi_*$, we can compute the CMB observables $n_s$ and $r$ as functions of the gauge boson mass, as shown in Fig.~\ref{nsrbosonnsr}. Similar to the results for scalar loops, the potential including corrections from gauge boson loops sets an upper limit on the gauge boson mass. The $n_s$ constraint from {\it Planck} yields the bound $m_V\lesssim 1.02\times 10^{16}$~GeV which can be compared to the limit from Eq.~(\ref{S2}) with $m_V = 5 g_5 v_\Sigma$. In contrast to the results from fermion loops (shown in Fig.~\ref{efffermion}), the value of $n_s$ decreases with increasing $m_V$, and it is the lower bound on $n_s$ from {\it Planck} data that sets the upper bound on $m_V$, as seen in the left panel of Fig.~\ref{nsrbosonnsr}.  In this case, the value of $n_s$ never reaches the range indicated by ACT.  Within this range, the tensor-to-scalar ratio stays well below the current experimental bound $0.036$. As seen in the right panel of Fig.~\ref{nsrbosonnsr}, $r$ takes values between 0.003 and 0.004, a range that can be explored by future observations.

\section{Summary}
\label{summ}

It is remarkable that there exist precision data that allow us to probe features of the early universe at its earliest moments during and after inflation.  The amplitude of the scalar perturbation spectrum sets the inflationary scale (see Eq.~(\ref{As2})) and the tilt of the power spectrum provides information on the shape of the scalar potential driving inflation (see Eq.~(\ref{nseq})). Hopefully, in the not too distant future additional information on the shape of the potential will be available from the measurements of the tensor perturbation spectrum, for which only an upper limit is currently available. 

It is also remarkable that one of the first models of inflation proposed (originally to solve the problem of the big bang singularity) \cite{Staro} remains in good agreement with CMB data. The Starobinsky potential (\ref{treestaro}) is very special, as it approximates true de Sitter ($R^2$ gravity) at large field values. Writing the theory in the Einstein frame, we obtain a tree-level scalar potential suitable for inflation. It is important to verify whether or not this effective description is valid when radiative corrections are considered. In a series of papers, we have explored this question. First \cite{Ellis:2025bzi}, we considered the radiative corrections induced by couplings of the inflaton to Standard Model fields. Such couplings are necessary for establishing a thermal background after inflation ends. Secondly\cite{Ellis:2026ceb}, we computed the radiative corrections induced by broken supersymmetry when the inflation model is derived from no-scale supergravity, which is closely related to Starobinsky's $R+R^2$ theory of gravity \cite{building,eno9,DLT}. These works were summarized in Sections \ref{radstar} and \ref{rcns} of this paper. 

One of the challenges in the construction of no-scale supergravity avatars of the Starobinsky model is attaining suitable reheating due to the suppressed couplings of the inflaton to matter \cite{Endo:2006xg,egno4,Ema:2024sit,Antoniadis:2026bys}. This problem and the possibility of reheating through the trace anomaly to gauge bosons were discussed in Section \ref{reheat}. While the couplings of the inflaton to Standard Model fields may provide acceptable reheating, they are generally so small that they do not perturb significantly the Starobinsky potential.  However, the same types of Lagrangian terms provide supergravity couplings to GUT fields as well. While these cannot be responsible for reheating, as decays to GUT-scale fields would be kinematically forbidden (decays through loops would be possible but suppressed if supersymmetry breaking is at the weak scale), they could affect the inflaton potential and CMB observables. This was discussed in Section \ref{rcreh} where new limits were obtained on the GUT $\mu$-terms such as $\mu_\Sigma$ for the SU(5) adjoint and its vev. Also, the anomalous inflaton coupling to GUT-scale gauge bosons allowed limits to be set on the GUT scale. 

Future CMB observations therefore offer the prospect of probing GUT radiative corrections and tightening constraints on the GUT mass scale.

\acknowledgments{ This review is dedicated to Ignatios Antoniadis on the occasion of his 70th birthday, in
recognition of his many contributions to theoretical physics a portion of which is represented here. The work of J.E. was supported by the United Kingdom STFC Grant ST/T000759/1.
The work of T.G. and K.A.O. was supported in part by DOE grant DE-SC0011842 at the University of Minnesota. The work of K.K. was supported in part by Niigata University Grant for the Enhancement of International Collaborative Research, 2025.
}

\reftitle{References}


\end{document}